\documentclass[showpacs,amsmath,amsart,twocolumn,showkeys,pra,superscriptaddress]{revtex4-1}
\usepackage{dcolumn}    
\usepackage{subfigure}
\usepackage{ifpdf}
\usepackage{amssymb,lineno,amsfonts}
\usepackage{graphicx}   
\usepackage{dcolumn}    
\usepackage{bm}         
\usepackage{bbm}
\usepackage{mathrsfs}
\usepackage{upgreek}
\usepackage{mathtools}
\usepackage{epstopdf}
\usepackage{setspace}
\usepackage{hyperref}
\usepackage{float}
\usepackage{natbib}
\usepackage[usenames,dvipsnames]{xcolor}
\usepackage[matrix,frame,arrow]{xypic}

\definecolor{med-blue}{RGB}{25,25,112}
\hypersetup{colorlinks, linkcolor={red},citecolor={blue}, urlcolor={MidnightBlue}}

\newcommand{\ket}[1]{\vert{#1}\rangle}

\newcommand{\outpr}[2]{\vert{#1}\rangle\langle{#2}\vert}

\newcommand{\expv}[1]{\langle{#1}\rangle}
\newcommand{\mprod}[3]{\langle{#1}\vert{#2}\vert{#3}\rangle}
\newcommand{\proj}[1]{\outpr{#1}{#1}}

\begin{document}
\title{Optimized dynamical protection of nonclassical correlation in a quantum algorithm}
\author{ V. S. Anjusha and T. S. Mahesh}
\email{mahesh.ts@iiserpune.ac.in}
\affiliation{Department of Physics and NMR Research Center,\\
Indian Institute of Science Education and Research, Pune 411008, India}

\begin{abstract}
{
A quantum memory interacts with its environment and loses information via decoherence as well as incoherence.  A robust quantum control that prepares, preserves, and manipulates nonclassical correlations even in the presence of environmental influence is of paramount importance in quantum information processing.  A well-known technique to suppress decoherence, namely Dynamical Decoupling (DD), consists of a sequence of rapid modulations in the system state in order to refocus the system-environment interactions.  In this work, we describe integrating DD with quantum gates using optimal control techniques in order to realize robust quantum gates protected against incoherence and decoherence.  
To investigate the protection of non-classical correlation, we study the evolution of quantum discord in Grover's search algorithm implemented with dynamically protected gates.   Using a two-qubit NMR system, we experimentally demonstrate a significant protection of quantum discord against incoherence and thereby retain a faithful probability profile of the marked state. While most of the DD sequences are based on spin-state inverting $\pi$-pulses, interestingly, we find that the protected gates integrated with non-inversion DD pulses may perform as effective or even better.  We also support the experimental results by analytical and numerical methods.
}
\end{abstract}

\keywords {decoherence, dynamical decoupling, nuclear magnetic resonance, quantum discord, optimal control}
\pacs{03.67.Lx, 03.67.Ac, 03.65.Wj, 03.65.Ta}
\maketitle

\section{Introduction}
Quantum information processing (QIP)
often provides classically inaccessible shortcuts to computational problems.
Quantum correlations such as quantum discord (QD) and quantum entanglement are regarded as  precious resources for QIP \cite{fanchini2017lectures,mahesh2017quantum,datta2008quantum}.  In this sense, it is important to preserve these resources during the process of computation.
A quantum register interacts with its environment and suffers loss of  information stored in it via decoherence.  In practice, another important source of information loss is via incoherence, which results from the spatial inhomogeneity in the control fields \cite{henry2007signatures}.
Therefore it is necessary to realize noise-free quantum controls which preserve quantum correlations. 
Dynamical decoupling (DD) involves modulating the system-environment interaction and thereby  suppress decoherence as well as incoherence \cite{viola1999dynamical,viola2002quantum}.
It usually consists of a sparse sequence of  instantaneous qubit-flips to systematically modulate the system-environment interaction.
Unlike the fault-tolerant schemes based on error-correction or decoherence-free subspaces, DD requires no additional resource in terms of ancillary registers \cite{souza2012robust,lidar2012review,suter2016colloquium}. 
Accordingly, DD has been widely studied and implemented experimentally \cite{alvarez2010performance,ajoy2011optimal,peng2011high,shim2012robust}.  Recently several experimental studies have been performed towards protecting quantum correlations in quantum memory
 \cite{katiyar2012evolution,singh2017experimentally}.
  More recently, combining DD and quantum gates had been studied theoretically \cite{khodjasteh2009dynamically} as well as experimentally \cite{dd_gte_nv_du,CR_dd,protectedgates_suter}.

In this work, we incorporate DD within the optimal control procedure, which not only avoids the manual slicing of the gate segments, but also naturally takes care of  DD pulse-errors.     In particular, we demonstrate realizing protected gates by combining DD and Gradient Ascent Pulse Engineering (GRAPE) protocol \cite{GRAPE}.   As a specific case, we study the experimental efficiency of protected gates implementing Grover's search algorithm in terms of preserving QD and of maintaining the probability of marked state.  

In sections II and III, we provide brief theoretical descriptions of DD-protected gates and QD respectively.  We shall describe the results of NMR experiments and numerical simulations in section IV.  Finally we conclude in section V.

\section{Dynamically protected gates}
\subsection{Theory}
Consider an N-spin system with an internal Hamiltonian $\cal{H}_S$ and a control Hamiltonian 
\begin{equation}
\mathcal{H}_C(t) =  \sum_{i=1}^N \Omega_{x}(t) I_{ix} + \Omega_{y}(t) I_{iy}
\end{equation}
where $\Omega_{x}(t)$ and $\Omega_{y}(t)$ represent the $x$ and $y$ components of the amplitude and phase modulated RF fields being applied on all the spins. 
Simultaneously, we also consider the system-environment interaction Hamiltonian ${\mathcal{H}}_{SE}(t)$, so that the total Hamiltonian is of the form
\begin{equation}
\mathcal{H}(t)=\mathcal{H}_S+\mathcal{H}_C(t)+\mathcal{H}_{SE}(t).
\end{equation}

While the system-environment interaction Hamiltonian ${\mathcal{H}}_{SE}$ that is responsible for the decoherence, is usually hard to characterize, it is still possible to minimize it's disruptive effect by controlled system modulations. The goal here is to achieve a quantum operation while protecting the quantum register against the environmental decoherence.  In the following, we first describe the procedure for integrating DD with Quantum Control (QC) to achieve a protected quantum operation.

\begin{figure}
	\includegraphics[trim=3cm 5cm 3.5cm 3cm,width=7.5cm]{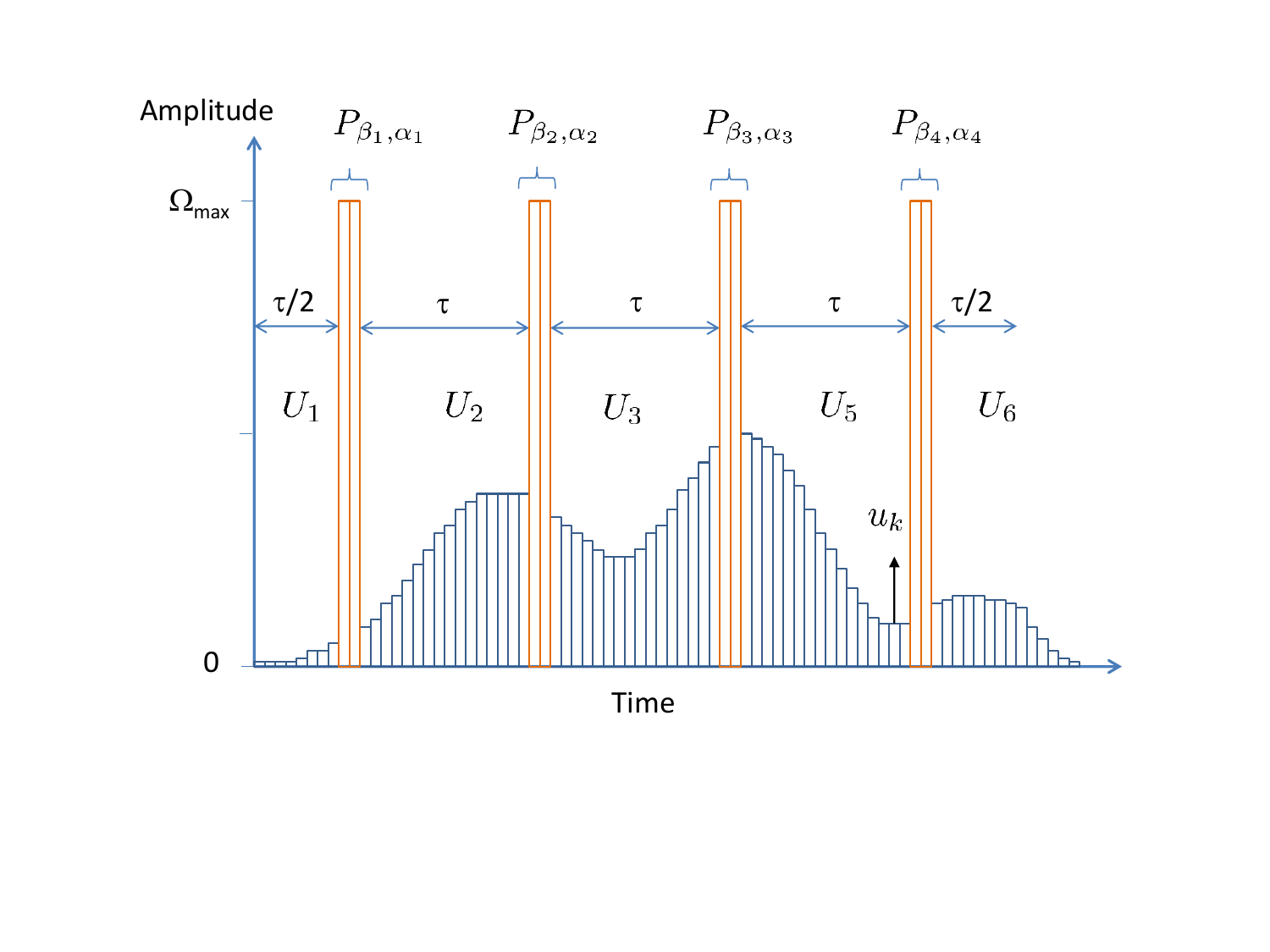}
	\caption{Protected quantum gate scheme. Certain segments are reserved for the full-amplitude ($\Omega_\mathrm{max}$) DD-pulses ($P_{\beta_j,\alpha_j}$) and other segments are subject to optimization to realize a given target unitary $U_T$.}
	\label{dd_scheme}
\end{figure}

Suppose a propagator $U_T$ is to be realized by the control fields $\{\Omega_{x}(t),\Omega_{y}(t)\}$.  In practice, time-discretized  (piecewise constant) amplitudes are used, i.e.,  during the $k$th segment $[(k-1)\Delta t,k\Delta t]$, the control parameters are  $\{\Omega_{x,k},\Omega_{y,k}\}$ (see Fig. \ref{dd_scheme}). 
The ideal closed-system unitary for the $k$th segment is
\begin{eqnarray}
u_k = \exp\left[-i({\mathcal H}_S+ {\mathcal H}_{C,k})\Delta t\right],
\end{eqnarray}
where
${\mathcal H}_{C,k} = \sum_i\Omega_{x,k}I_{ix}+
\Omega_{y,k}I_{iy}$.
The control propagator for a sequence of n-segments is $U = u_k u_{k-1} \cdots u_1$. 

In the presence of environmental interaction however, the actual open-system propagator for the $k$-th segment becomes
\begin{eqnarray}
u_k^\mathrm{op} = \exp\left[-i({\mathcal H}_S+ {\mathcal H}_{C,k} + {\mathcal H}_{SE})\Delta t\right],
\end{eqnarray}
which acts on the joint system-environment state.  Subsequent tracing out of the environmental part results in decoherence of the quantum system.  
DD pulses modulate the system-environment evolution and thereby suppress decoherence.

The propagator for a DD pulse with flip-angle $\beta$ and phase $\alpha$ is given by $P_{\beta,\alpha} = \exp(-i\beta S_\alpha)$, where $S_\alpha = \sum_i I_{ix} \cos \alpha +   I_{iy} \sin\alpha$.
The overall propagator for a protected sequence consisting of $M$ DD pulses sandwiched between control propagators is (see Fig. \ref{dd_scheme}),
\begin{eqnarray}
U_P = U_{M+1} \prod_{j=1}^M P_{\beta_j,\alpha_j} U_j.
\end{eqnarray}
  
Using the toggling-frame picture \cite{haeberlen2012high,protectedgates_suter}, we may rewrite the above in the form,
\begin{eqnarray}
U_P = U_{M+1} \prod_{j=1}^M \tilde{U}_j,
\end{eqnarray}
where the toggling frame unitaries $\tilde{U}_j = T_{j}^\dagger U_j T_{j}$, and $T_j = P_{\beta_{j-1},\alpha_{j-1}}P_{\beta_{j-2},\alpha_{j-2}} \cdots P_{\beta_{1},\alpha_{1}}$ and $T_1 = T_{M+1} = \mathbbm{1}$.  
Given a target propagator $U_T$, we optimize the control amplitudes which maximize the fidelity
\begin{eqnarray}
F(U_P,U_T) = \mathrm{Tr}\left[U_T^\dagger U_P\right]/\mathrm{Tr}\left[U_T^\dagger U_T\right].
\label{fidelity}
\end{eqnarray}

Unlike the previous works on the protected gates \cite{protectedgates_suter,CR_dd}, where $\beta = \pi$ was used, in this work we generalize to variable DD flip angles. We also integrate periodic, full-amplitude, DD pulses in the optimal control procedure itself, by pre-assigning segments and freezing their amplitudes and phases.  The main advantage of this method is that the DD pulses along with their imperfections, like additional spin-flips, RF inhomogeneity, and offset errors, are accounted and corrections are incorporated by the control segments in order to achieve a robust sequence for implementing the target propagator. Moreover, there is no need for the manual slicing of control pulses to introduce DD pulses.
In the following, we shall consider the dynamical protection of a single qubit and illustrate how it can be achieved by DD pulses with a range of flip-angles.

\subsection{A simple model: \\Protected NOT gate on a single qubit}
\begin{figure}
	\subfigure[]{\includegraphics[trim=0cm 0.1cm 0cm 0.2cm,width=6cm,clip=]{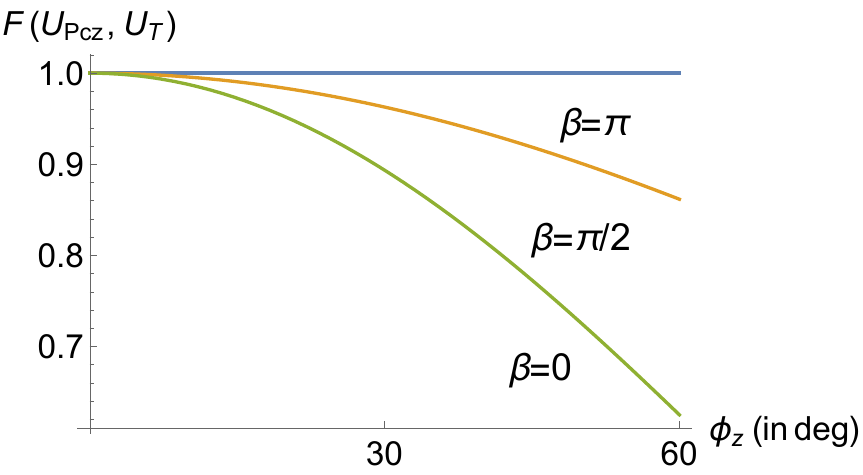}} \\
	\subfigure[ ]{\includegraphics[trim=4cm 7cm 5cm 8cm,width=7cm]{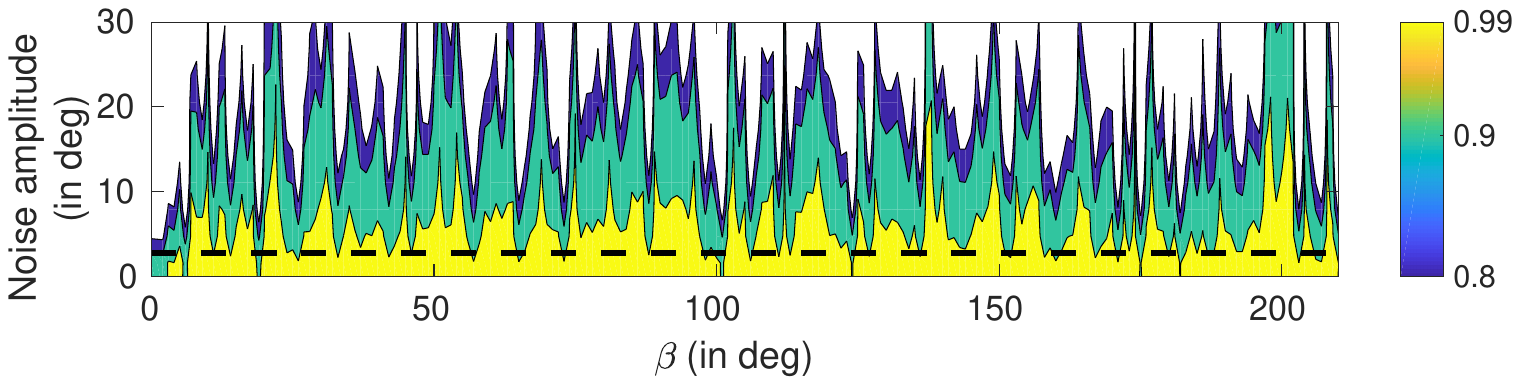}}
	\caption{(a) Fidelity of the protected NOT gate
	 versus offset error (in deg) corresponding to the DD sequences $U_{Pcz} = U_{cz} P_{\beta,x} U_{cz}$. (b) Fidelity of the protected NOT gate $U_{cz}^{(y)} P_{\beta,y} U_{cz}^{(y)} U_{cz}^{(x)} P_{\beta,x} U^{(x)}_{cz}$ versus DD angle $\beta$ and amplitude of the random phase-rotations.  For the unprotected NOT gate, the fidelity drops below 0.9 for the noise amplitude $\phi_z > 3^\circ$ as indicated by the dashed line.}
	\label{singlequbitdd}
\end{figure}

For a simple demonstration of dynamical protection, we choose the NOT gate, $U_T = \exp(-i \pi I_x)$ on a single qubit.  We consider a DD pulse $P_{\beta,x} = \exp(-i \beta I_x)$ sandwiched between two identical effective control propagators $U = \exp(-i \pi/2~ I_x)$, such that the total propagator $U_P = U P U$.  Here the DD pulse has not been accounted by the control propagators, and therefore the fidelity $F(U_P,U_T) = \cos(\beta/2)$ drops with the flip-angle.  In order to account for the DD pulse, we introduce a correction in the control propagator such that $U_c = \exp\{-i (\pi/2-c)I_x\}$, where $c$ is the correction factor.  In this case,  fidelity of the target with the net corrected propagator $U_{Pc} = U_c P_{\beta,x} U_c$ becomes $F(U_{Pc},U_T) = \cos(\beta/2-c)$.  Thus setting $c=\beta/2$ leads to unit fidelity for arbitrary DD flip-angles.  We now consider an undesired offset-error $\phi_z$ such that the DD-corrected control propagators take the form $U_{cz} = \exp[-i \{(\pi/2-\beta/2)I_x + \phi_z I_z\}]$, then the fidelity of the operation $U_{Pcz} = U_{cz} P_{\beta,x} U_{cz}$ becomes
\begin{eqnarray}
F(U_{Pcz},U_T) &=&
\left( \frac{4\phi_z^2}{\gamma^2}+ \frac{\eta^2}{\gamma^2} \cos\frac{\gamma}{2} \right)  \sin\frac{\beta}{2}
+
 \frac{\eta}{\gamma} \sin\frac{\gamma}{2} \cos\frac{\beta}{2}, \nonumber
\end{eqnarray}
where $\gamma^2 = \eta^2+4\phi_z^2$ and $\eta = \pi-\beta$.  Clearly $F(U_{Pcz},U_T) = 1$ for $\beta = \pi$.  However, for small values of $\phi_z$, $F(U_{Pcz},U_T) \approx 1$ for all values of $\beta$.   Without the DD pulse, the fidelity of the unprotected operation $U_{z}^2 = \exp\{-2i(\pi/2~ Ix+\phi_z I_z)\}$ is $F(U_z^2,U_T) =  \frac{\pi}{\gamma} \sin\frac{\gamma}{2}$, which starts from unity for $\phi_z=0$, but drops as $\phi_z$ starts increasing.
Fig. \ref{singlequbitdd}(a) displays the relative performances of these DD sequences with respect to the offset-error $\phi_z$.  For this simple model, it turns out that while $\beta = \pi$ works the best, even $\beta=\pi/2$ can show significant protection.

We now consider a more general case of a XY-DD protected NOT gate
$U_{cz}^{(y)} P_{\beta,y} U_{cz}^{(y)} U_{cz}^{(x)} P_{\beta,x} U^{(x)}_{cz}$, where 
$$U_{cz}^{(x/y)} = \prod_{j=1}^n \exp\left\{-i  \left(\frac{\pi}{4n} - c_{x/y}I_x - d_{x/y}I_y + \phi_z \eta_j \right)\right\}$$ subjected to a random noise in the range  $[-\phi_z\eta_j,+\phi_z\eta_j]$ with a variable amplitude $\phi_z$.  We have studied the above gate with $n=100$ segments and numerically optimized the correction factors $c_{x/y}I_x$ and $d_{x/y}$.  During the above optimization, noise term $\phi_z\eta_j$ need not be considered.  The contour plot of the fidelity of such an XY-DD protected NOT gate is shown in Fig. \ref{singlequbitdd}(b).  For the unprotected NOT gate, the fidelity drops below 0.9 for $\phi_z > 3^\circ$ as  shown by the dashed line.  However in this region, fidelity of the protected NOT gate is above 0.99 for a wide range of DD angles.  These models indicate that unlike the coherence storage schemes where $\beta = \pi$ is usually considered, protected quantum gates can also be constructed out of $\beta \neq \pi$.

Before we discuss the experimental and numerical studies, we first briefly review quantum discord and its evolution during different stages of Grover's algorithm. 

 \begin{figure}
 	\includegraphics[trim=6.7cm 13cm 6.8cm 12cm,width=5cm]{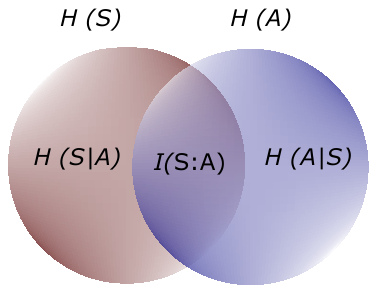}
 	\caption{Venn diagram representing the total information $H(S,A)$ of a bipartite system. $ H(A)$ and $H(S)$ are individual information of the system A and S respectively, $I(S:A)$ is the mutual information and $H(S|A)$ and $H(A|S)$ are conditional entropies.  }
 	\label{QD}
 \end{figure}

\section{Quantum Discord (QD)}
Quantum entanglement is considered as an important resource in QIP \cite{raussendorf2001one,bennett2000quantum}.  Even when the entanglement is absent, a bipartite quantum system $(S,A)$ may possess another useful type of quantum correlation known as quantum discord (QD) \cite{QD_zurek}. QD is quantified by the minimal mismatch between the mutual information obtained in two classically equivalent, but quantum mechanically distinct ways. As can be seen from Fig. \ref{QD}, the mutual information can be expressed either as 
\begin{equation}
I(S:A)=H(A)+H(S)-H(S,A),
\end{equation}
or as
\begin{equation}
J(S:A)=H(S)-H(S|A).
\end{equation}
In classical information, 
\begin{eqnarray}
H(X) = -\sum_x p_x\log_2 p_x
\end{eqnarray}
is the Shannon entropy obtained using the probabilities $p_x$ of $x$th outcome.  Similarly, the joint entropy
\begin{eqnarray}
H(S,A) = -\sum_{a,s} p(s,a) \log_2 p(s,a),
\end{eqnarray}
is obtained using the joint probabilities $p(s,a)$ and the conditional entropy 
\begin{eqnarray}
H(S|A) = -\sum_{a,s} p_a p(s|a) \log_2 p(s|a),
\end{eqnarray}
is obtained using the conditional probability  $p(s|a)$, which is the probability of occurrence of outcome $s$ given that the outcome $a$ has occurred.

In quantum information, we replace the Shannon entropy with von Neumann entropy, i.e.,
\begin{eqnarray}
H(\rho_X) = -\sum_x \lambda_x\log_2(\lambda_x),
\end{eqnarray}
where $\lambda_x$ are the eigenvalues of the density matrix $\rho_X$.  The joint von Neumann entropy $H(S,A)$ is evaluated using the eigenvalues $\lambda_{s,a}$ of the full density matrix $\rho_{S,A}$.  On the other hand, the conditional entropy $H_{_\Pi}(S|A)$ is evaluated using the eigenvalues $\lambda_{s|a_\Pi}$ of the post-measurement density matrix $\rho_{S|A_\Pi}$ after carrying out a measurement along a basis $\{\Pi_i^a\}$. 

While the two forms $I(S:A)$ and $J(S:A)$ of mutual information are identical in the classical case, the requirement of measurement for the latter form makes them different in the quantum case.
This difference, when minimized over the entire set of orthormal measurement bases $\{\Pi\}$, is often nonvanishing, and is attributed to a form of quantum correlation termed as quantum discord (QD) \cite{QD_zurek,PhysRevA.83.012327}, i.e.,
\begin{equation}
D(S|A)= I(S:A)-\max_{\Pi}J(S:A).
\end{equation}
Since QD is also regarded as an important resource for quantum computing, it is necessary to understand its evolution during quantum algorithms.  In the following we discuss the evolution of QD in Grover's search algorithm.

\begin{figure}
	\includegraphics[trim=1cm 0cm 0cm 0cm,width=8.9cm]{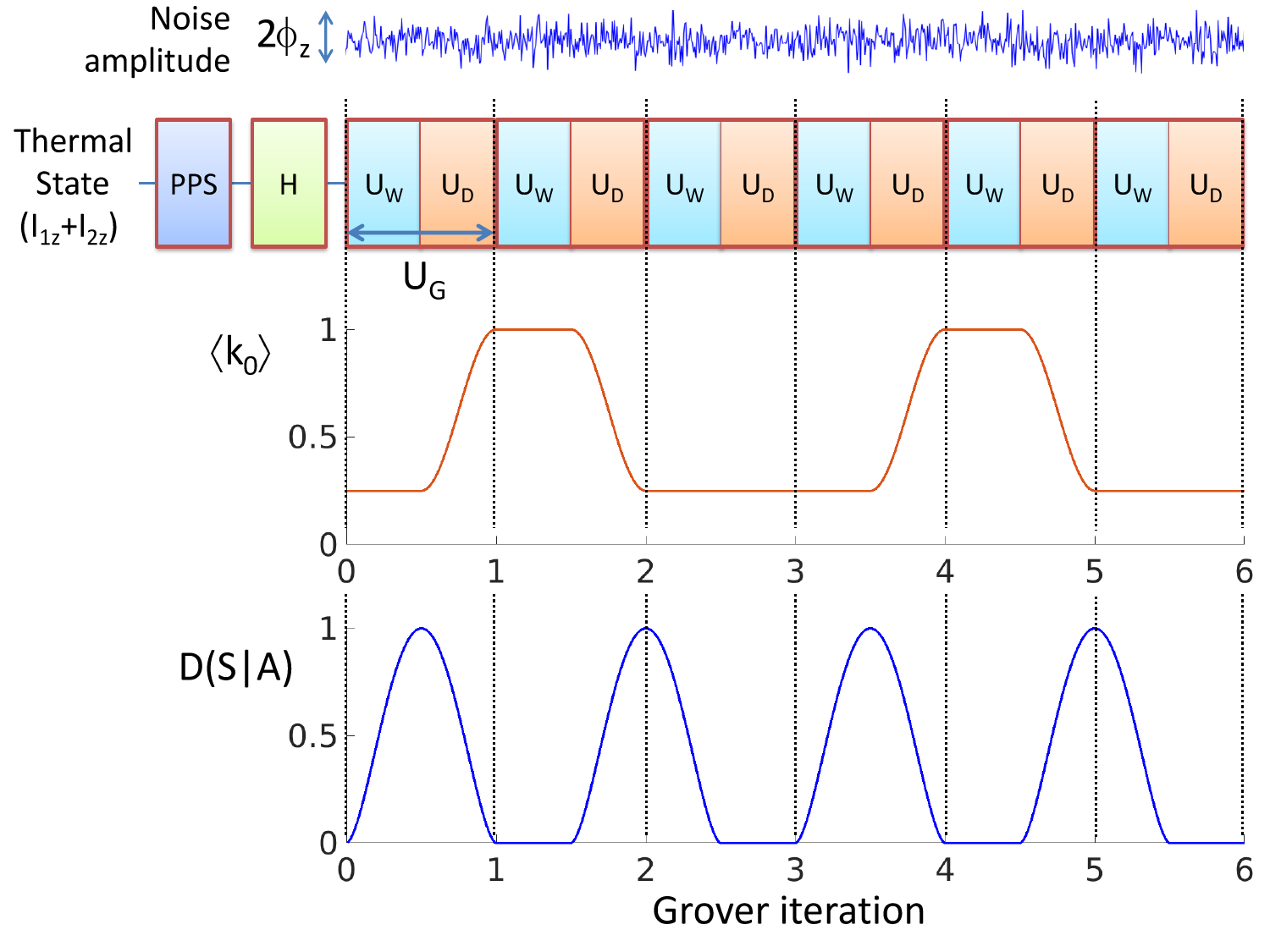}
	\caption{Various experimental stages of Grover's search algorithm for up to six iterations subjected to a random noise (top trace).  It begins with a thermal initial state, followed by preparation of pseudopure state (PPS),  Hadamard operator (H), and application of Grover's iterates ($U_G$) consisting of oracle ($U_W$) and diffusion ($U_D$) operators.
		The middle and bottom traces show ideal evolutions of probability of marked state $\expv{k_0}$ and QD respectively.}
	\label{grover}
\end{figure}

\begin{figure*}
	\includegraphics[clip=,
	trim=2.7cm 4.5cm  3cm 4.5cm, width=18cm]{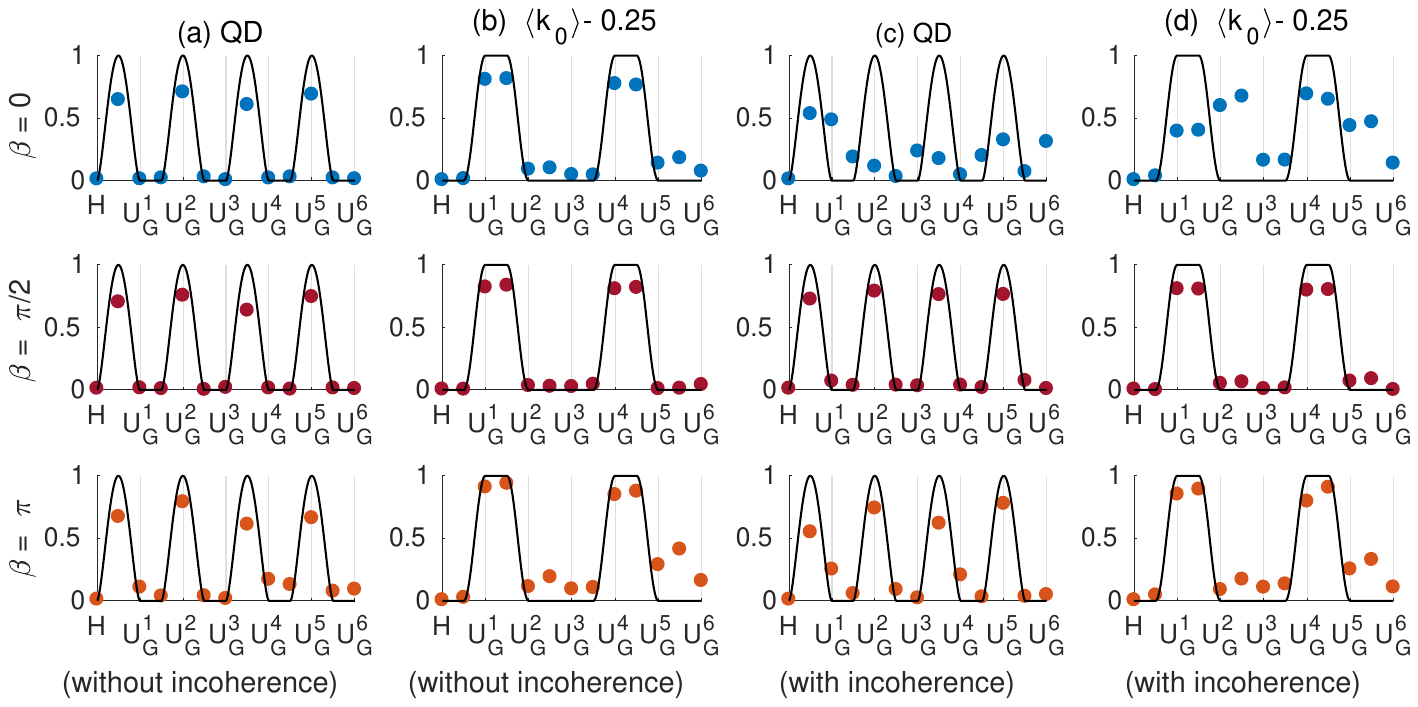}
	\caption{Idealized estimations (solid lines) and experimental (filled circles) QD (in units of $\epsilon^2/\ln2$) (a, c) and probability $\expv{k_0}-0.25$ (in units of $\epsilon$;  constant 0.25 is due to the traceless deviation matrix experimentally estimated) of the marked state  (b, d) under Grover iterates with XY DD-protections without (a, b) and with (c, d) additional incoherence. The top trace ($\beta = 0$) corresponds to  unprotected Grover iterates. The middle ($\beta = \pi/2$) and the bottom ($\beta = \pi$) traces correspond to XY DD protected Grover iterates. }
	\label{noise}
\end{figure*}

\subsection*{QD in Grover's algorithm} 
Grover’s algorithm identifies a marked item in an unsorted database of $N$ items in $O(\sqrt{N})$ iterations, while a classical algorithm needs $O(N)$ iterations on an average.  The algorithm starts with initializing a quantum register into an uniform superposition
\begin{eqnarray}
\ket{\psi_0} = \frac{1}{\sqrt{N}}\sum_{k=0}^{N-1} \ket{k}.
\end{eqnarray}
It then uses an oracle 
\begin{eqnarray}
U_W = \sum_{k=0}^{N-1} (-1)^{\delta_{k,k_0}} \proj{k}
\end{eqnarray}
that flips a marked state $\ket{k_0}$, and
a diffusion operator
\begin{eqnarray}
U_D = 2\proj{\psi_0} - \mathbbm{1},
\end{eqnarray}
which inverts each of the basis states about the mean.  Together these two operations constitute a Grover's iterate, i.e., $U_G = U_D U_W$, that amplifies the marked state amplitude.   

The top trace of Fig. \ref{grover} displays various stages in the Grover's algorithm up to six iterations.  The middle trace displays the periodic evolution of the probability $\expv{k_0}$ of the marked state $\ket{k_0} = \ket{01}$.  
The bottom trace displays the corresponding evolution of QD over six iterations.  As expected, QD vanishes whenever the system reaches the marked state - which is a classical state.  QD vanishes also for uncorrelated states of the form $\rho_S\otimes \rho_A$ not involving mutual interaction between $S$ and $A$.  
The question that we ask is, how does QD evolve in a noisy channel implementing Grover's algorithm, and how well it can be protected by interleaving the Grover's iterates with DD.  In the following section, we explore the answer to the above question by NMR based experiments.

\section{Experiments}
In our NMR experiments we utilize the two spin-1/2 proton nuclei of Cytosine dissolved in deuterated dimethylsulfoxide (DMSO-D$_6$).   
All the experiments are carried out on a Bruker 500 MHz NMR spectrometer at an ambient temperature of 300 K. The resonance offsets of the two protons are $436$ Hz and $-436$ Hz, while the scalar coupling constant $J = 7.0$ Hz.
Starting from the thermal equilibrium state $I_{Sz}+I_{Az}$, we use the spatial averaging technique \cite{corypps} to prepare  the pseudopure state (PPS) $(1-\epsilon)\mathbbm{1}/4+\epsilon\proj{00}$, where $\epsilon\sim 10^{-5}$ is the purity factor typically present in NMR systems.   

We generated the DD-protected oracle operator $U_W$ corresponding to the marked state $\ket{k_0} = \ket{01}$ and the diffusion operator $U_D$ by incorporating DD pulses into the GRAPE optimal control technique \cite{GRAPE,bhole2017rapid} as described in section II.  
Each GRAPE segment was of duration 5.1 $\upmu$s.  The fidelities of these GRAPE pulses, each about 75 ms long, were above 0.99 after averaging over 10\% RF inhomogeneity distribution.  A full power DD pulse was introduced in between every 1000 segments.  The phase $\alpha$ of the DD pulses was alternated between $x$ and $y$.

To estimate QD, we need to obtain the density matrix at various stages of Grover's algorithm (see Fig. \ref{grover}). The density matrix is generally obtained using quantum state tomography (QST) which involves a set of independent experiments (on identically prepared states) each measuring a particular set of observables.  The expectation values are then obtained by measuring the signal intensities.  We have adopted a QST procedure that results in only absorptive spectral lines which precisely quantify the expectation values without requiring any further numerical processing (see Appendix).

After obtaining the experimental density matrix $\rho_\mathrm{exp}$, we estimated QD via the optimal set of measurement bases proposed by Lu et al. \cite{PhysRevA.83.012327}. Similarly, we also estimated the probability $\expv{k_0}= \mprod{k_0}{\rho_\mathrm{exp}} {k_0}$ of the marked state for $\ket{k_0} = \ket{01}$.
Fig. \ref{noise} (a) displays experimentally estimated QD values (in units of $\epsilon^2/\ln2$) at various stages of the Grover's algorithm for up to six iterations,
with unprotected (top trace) as well as for XY DD-protected Grover's iterates with $\beta = \pi/2$ (middle trace) and $\beta = \pi$ (bottom trace).
Fig. \ref{noise} (b) displays the corresponding probability of the marked state $\ket{01}$.  
Thus, in the absence of incoherence, we observed that the experimental values are in reasonably good agreement with the idealized predictions represented by solid lines (also described in Fig. \ref{grover}) indicating high-quality of quantum controls as well as low inherent noise in the NMR system.

To further investigate the power of DD protection, we deliberately introduced incoherence in the form of linear static-field inhomogeneity along the z-axis by using approximately 5 mG/cm pulsed-field-gradient (PFG).  It introduces a differential offset of $\pm 10$ Hz across the sample volume corresponding to a phase shift $\phi_z \sim 0.1^\circ$ during a single GRAPE segment of duration 5.1 $\upmu$s. The total duration for six Grover-iterations was about 0.9 second, and the translational diffusion of the molecules over this period further introduces randomness in the overall dephasing.  
The top trace of Fig. \ref{noise} (c,d) display respectively the experimentally measured QD values and probabilities with unprotected Grover iterates in the presence of such an incoherence.  Clearly, the experimental data without DD protection show little correlation with the ideal trajectories.  The other traces of Fig. \ref{noise} (c,d) display the experimental data under XY DD-schemes with $\beta = \pi/2$ and $\beta = \pi$ as indicated.  The best protection was achieved by the XY DD sequence consisting of $\pi/2$ pulses.

\begin{figure}
	\includegraphics[clip=,trim=2.6cm 2.6cm 4cm 2cm,width=8.5cm]{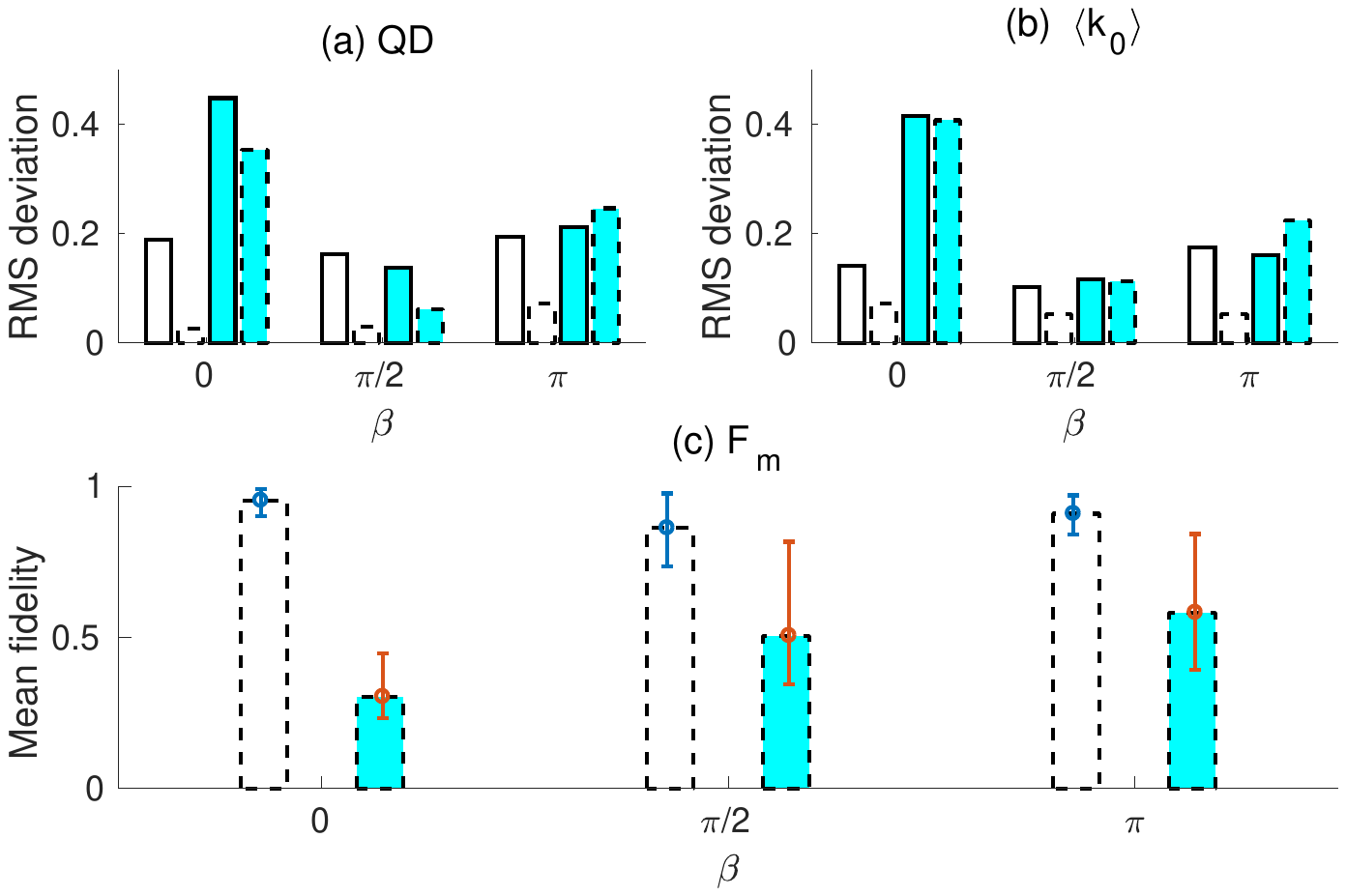}
	\caption{Average root-mean-square deviation (bars with solid edges) between the experimental data and ideal theoretical values (shown in Fig. \ref{noise}) for QD (in units of $\epsilon^2/\ln2$) (a) and the probability $\expv{k_0}$ (in units of $\epsilon$) (b) of marked state  versus DD flip angle $\beta$  without incoherence (open bars) and with incoherence (filled bars). The corresponding numerical simulations are shown by bars with dashed edges.  (c) Numerically estimated  fidelities of $U_G$ averaged over six iterations without (open bars) and with (filled bars) incoherence.  Here errorbars indicate variations over six iterations.}
	\label{rms}
\end{figure}

The bars with solid edges in Fig. \ref{rms} show the average root-mean-square (RMS) deviation of the experimental QD (a) and probabilities $\expv{k_0}$ (b) from the idealized theoretical values displayed in Fig. \ref{noise} for various DD flip angles without (open bars) and with (filled bars) incoherence.  In general, RMS deviations are enhanced by incoherence, but suppressed by DD. Interestingly, the DD protection with $\beta=\pi/2$ showed the best performance in suppressing the effects of incoherence.  To further strengthen this claim, we simulated RMS deviations using GRAPE pulses corresponding to $U_{PG}$ for six iterations  with 20\% inhomogeneous RF fields and incoherent fields ranging from $-10$ Hz to $+10$ Hz.  The results are shown by bars with dashed edges in Fig. \ref{rms} (a,b).  The simulations also support the experimental finding that $\beta = \pi/2$ DD protection has a superior performance.
We also estimated the mean fidelities $F_m = \frac{1}{6}\sum_{j=1}^6 F(U_{PG}^j,U_{G}^j)$  (see Eq. \ref{fidelity}) as displayed in Fig. \ref{rms} (c). All the mean fidelities have good values in the absence of incoherence again indicating a good control.  However in the presence of incoherence,
while the unprotected iterate completely fails, both $\beta = \pi/2$ and $\beta = \pi$ cases show relatively better performances.

\begin{figure}
	\includegraphics[clip=,trim=1.3cm 6.4cm 2.3cm 6.2cm,width=8.5cm]{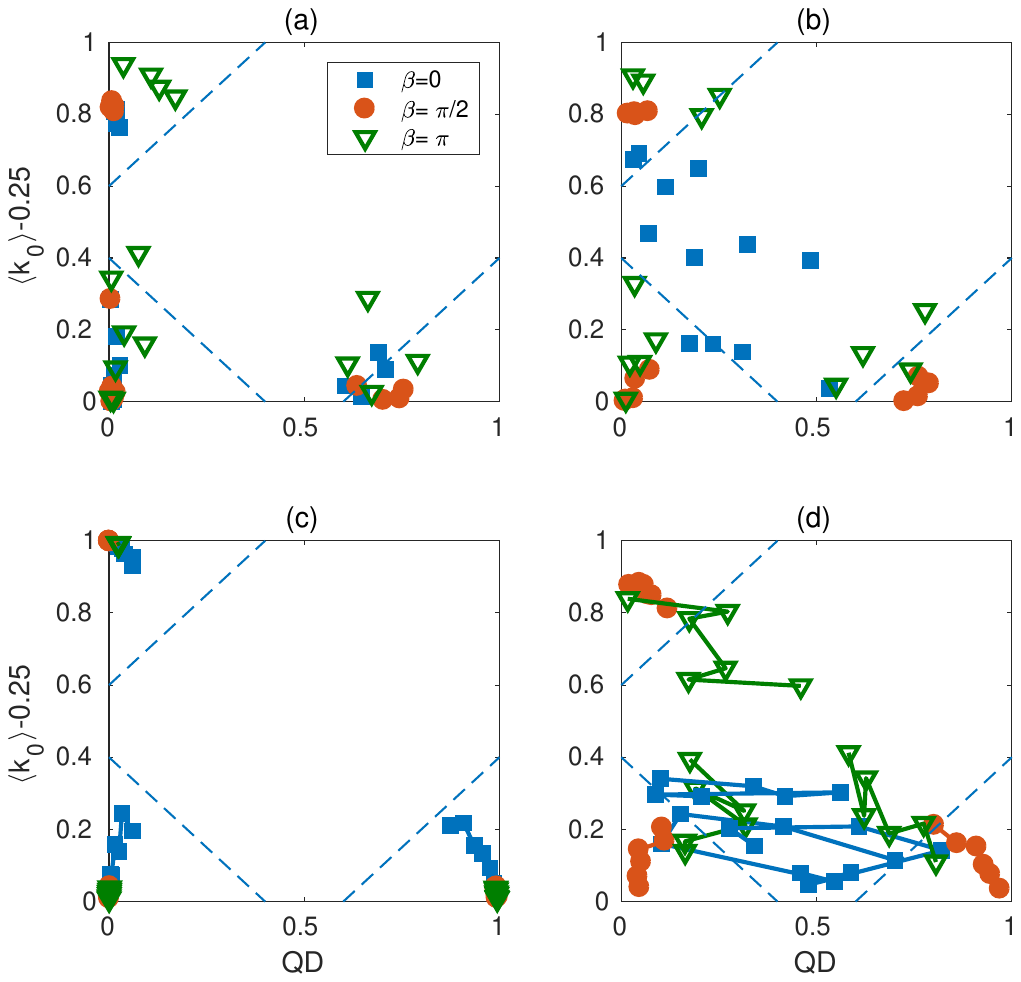}
	\caption{Experimental values of QD for six Grover iterations plotted against the probability $\expv{k_0}-0.25$ without (a) and with (b) incoherence. Simulated values of QD versus probability for 10 Grover iterations without (c) and with (d) incoherence. In all the cases, QD values are in units of $\epsilon^2/\ln 2$ and probability values are in units of $\epsilon$.}
	\label{k0vsqd}
\end{figure}

We now try to understand the correlation between  probability of the marked state and QD. Fig. \ref{k0vsqd} displays the experimental QD values versus the probabilities without (a) and with (b) incoherence.  Ideally, the initial superposition state has no QD, but has same probability of 0.25 for all the four basis states, which corresponds to the bottom left corner as indicated by the dashed line.  The intermediate state after the oracle $U_W$ takes the superposition to a highly correlated state with a maximum QD, but still without any higher probability for the marked state, which corresponds to the bottom-right corner.  Diffusion operator then amplifies the probability of the marked state, which is a classical state with no QD, and therefore corresponds to the top-left corner.
In the case of no incoherence, most of the data points are localized to the expected corners.  However, in the presence of incoherence, while there is a significant dispersion of data points belonging to the unprotected gates ($\beta = 0$), those belonging to protected gates ($\beta \neq 0$) still maintain localization. Thus DD schemes preserve quantum correlations and thereby assure success of the algorithm.  Here again $\beta=\pi/2$ shows the best performance.  Figures \ref{k0vsqd} (c,d) display the simulated values of probability versus QD without (c) and with (d) incoherent fields (ranging from $-10$ Hz to $+10$ Hz) for 10 Grover iterations.  In the absence of incoherence there is a reasonable localization for all the three values of $\beta$.  However, with incoherence the unprotected Grover iterations completely fail since the maximum probability of the marked state remained less than 0.4 throughout the iterations.  The data points corresponding to protected Grover iterates with $\beta = \pi$ are significantly dispersed, but still show signatures of localization, while those with $\beta = \pi/2$ show the best localization consistent with the experimental results.

\section{Conclusions}
Quantum correlations such as quantum discord and quantum entanglement form resources that fuels  quantum information processors.
In this work, we experimentally studied the evolution of quantum discord as well as the probability of a marked state over six iterations of Grover's  quantum search algorithm on an ensemble of spin-1/2 nuclear pairs using nuclear magnetic resonance methods.  
Unlike the earlier works on protected quantum gates which considered only $\pi$ pulses, we have generalized to DD-pulses with variable flip angles.
 Further, we have integrated the dynamical protection into GRAPE optimal control protocol by pre-assigning the positions of dynamical flips.  In this way,  protected quantum gates are robust by construction against the external noise. Similar protocol can also be incorporated in other optimal control techniques such as Bang-Bang \cite{BB_bhole}, Krotov \cite{maximov2008optimal}, etc.  
While the protected gates performed generally better, to investigate the extent of protection, we introduced an additional incoherent noise in the form of a pulsed field gradient.  In this case, we observed a significant benefit of dynamically protected gates.  Interestingly, we found that the dynamical protection with $\pi/2$ flip-angles to  outperform those with $\pi$.  
We have supported our experimental findings with numerical simulations and
also provided simple single-qubit models to explain this observation.  
Although, DD with non-inversion pulses might appear counter intuitive at the outset, $\pi/2$-rotation based modulation schemes such as WAHUHA \cite{gerstein1981high} and solid-echo sequences \cite{haeberlen2012high} have long been used in NMR spectroscopy.
We believe that our study will be useful in understanding dynamical protections and thereby designing robust quantum controls.

\section*{Acknowledgment}
We gratefully acknowledge Gaurav Bhole for suggestions in coding quantum controls.
We also acknowledge useful discussions with  Sudheer Kumar and Deepak Khurana. This work was partly supported by DST/SJF/PSA-03/2012-13 and CSIR 03(1345)/16/EMR-II.

\bibliographystyle{apsrev4-1}
\bibliography{bibqd}
\section*{appendix}
\subsection*{Pure-phase Quantum State Tomography}
In the conventional scheme (like the ones mentioned in \cite{nielsenchuang}, \cite{roy2010density},  \cite{PhysRevA.87.062317}), one obtains absorptive, or dispersive, or even mixed-phase spectral lines which are often hard to quantify.  A pure-phase tomography has been designed to obtain only absorptive spectral lines which are far easier and precise to quantify.  In the case of a two-qubit homonuclear system involves a set of six experiments:\\
$\begin{array}{cl}
(i) & \triangleright \equiv G_z - 60_{90}-\tau  \\
(ii) & 75_{-180}  - \triangleright \\
(iii) & 75_{105}  - \triangleright \\
(iv) &  1/2J - 75_{30}  -\triangleright \\
(v) & 5/12J - 75_{105}  -\triangleright \\
(vi)&  75_{195} - 5/12J - 75_{15}  - \triangleright.
\end{array}$ \\
Here $G_z$ is the pulsed-field gradient to destroy the coherences, $\tau$ is the delay optimized to suppress the homonuclear zero-quantum coherence, and $J$ is the indirect spin-spin coupling constant.  Four absorptive transitions obtained after each of the above experiments are integrated and the density matrix is estimated using the constraint-matrix procedure described in \cite{roy2010density,PhysRevA.87.062317}.

\end{document}